\documentclass{article}
\usepackage[utf8]{inputenc}
\usepackage{natbib}
\usepackage{authblk}
\usepackage{setspace}
\usepackage{xcolor}
\usepackage[margin=1.25in]{geometry}
\usepackage{graphicx}
\usepackage{hyperref}

\usepackage{times}
\usepackage{latexsym}

\usepackage{subcaption}
\usepackage{xspace}

\graphicspath{ {./figures/} }
\usepackage{subcaption}
\usepackage{amsmath}
\usepackage{caption}

\newcommand{\treated}{{\textcolor{blue}{\textsc{IdentityAdded}}}\xspace}
\newcommand{\untreated}{{\textcolor{olive}{\textsc{NotAdded}}}\xspace}
\newcommand{\control}{{\textcolor{red}{\textsc{Control}}}\xspace}
\newcommand{\alwayspositive}{{\textcolor{teal}{\textsc{AlwaysPositive}}}\xspace}
\newcommand{\alwaysnegative}{{\textcolor{violet}{\textsc{AlwaysNegative}}}\xspace}

\title{Profile Update: The Effects of Identity Disclosure on Network Connections and Language}

\author[1]{Minje Choi}
\author[1,2,3]{Daniel M. Romero}
\author[1,2]{David Jurgens}

\affil[1]{School of Information, University of Michigan}
\affil[2]{Computer Science and Engineering Division, University of Michigan}
\affil[3]{Center for the Study of Complex Systems, University of Michigan}
\affil[*]{Address correspondence to: minje@umich.edu}

\date{}

\begin{document}

\maketitle

\begin{abstract}
Our social identities determine how we interact and engage with the world surrounding us. 
In online settings, individuals can make these identities explicit by including them in their public biography, possibly signaling a change to what is important to them and how they should be viewed.
Here, we perform the first large-scale study on Twitter that examines behavioral changes following identity signal addition on Twitter profiles. Combining social networks with NLP and quasi-experimental analyses, we discover that after disclosing an identity on their profiles, users (1) generate more tweets containing language that aligns with their identity and (2) connect more to same-identity users. We also examine whether adding an identity signal increases the number of offensive replies and find that (3) the combined effect of disclosing identity via both tweets and profiles is associated with a reduced number of offensive replies from others. 
\end{abstract}


\section{Introduction}
Our \textit{social identities} such as age, gender, or occupation play a crucial role in shaping how we express thoughts and opinions through language, and in turn, how others interact with us. In social media platforms such as Twitter, the identity that one chooses to associate oneself with can influence behaviors such as the topics one engages with or the ties one forms. One can choose to explicitly disclose their identity through various means, but the effects and consequences of such actions are largely unknown. In this paper, we perform a large-scale study to understand how explicitly disclosing social identities leads to changes in the interactions of one's social network.


\begin{figure}[t!]
    \centering
    \includegraphics[width=.8\textwidth]{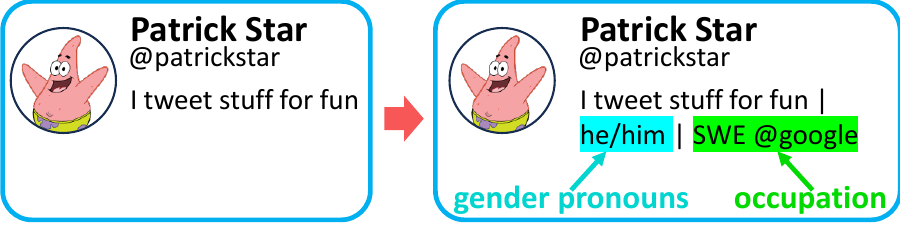}
    \caption{\small Identity disclosure in Twitter profiles}
    \label{fig:overview}
\end{figure}

Identity disclosure and management is an essential part of online behavior~\citep{joinson2010disclosure,pavalanathan2015identity}, as individuals navigate what aspects of themselves are salient to others. In more public platforms like Twitter, individuals must weight how to present themselves based on the mix of audiences who may find them~\citep{marwick2011tweet,bazarova2014disclosure,duguay2016he}.
People may explicitly express social identities in social media by including phrases related to the identity in \textit{profile descriptions}, as shown in Figure~\ref{fig:overview}. Profile descriptions, similar to posts, also contain rich textual features associated with the user's social identity~\citep{li2014twitterprofile,priante2016whoami,wilson2020twitterprofile,wang2019m3}.
Crucially, these profiles are not static: Individuals add and remove identity markers from their bios to emphasize new or specific aspects of themselves, such as political affiliations~\citep{jones2021twitterbio} or gender pronouns~\citep{tucker2023pronoun,jiang2022pronouns}.

Disclosed social identities can affect how they are perceived and targeted by other users. Prior studies have drawn connections between the disclosure of identities---especially marginalized or minority identities---and identity-based hate or cyberbullying, therefore hindering people from fully expressing themselves and sometimes even forcing them to hide identities online~\citep{haimson2015disclosure,jhaver2018online}. However, not all identities are marginalized and the potential varied outcomes for identity disclosure are yet to be quantified.

To understand the effects of identity disclosure, we conduct a large-scale quasi-experimental study on hundreds of thousands of users who updated their profiles to disclose a particular social identity. We observe that while overall tweet activity levels remain stable post-disclosure, their tweets contain significantly higher volumes of identity-relevant language, which we further dissect into topic and style properties. We demonstrate that this disclosure is also associated with social network changes: users actively engage more with similar-identity individuals following disclosure. Finally, we examine the number of offensive replies received from others during pre- and post-disclosure periods, where we show that contrary to existing studies~\citep{chan2022gender,meyer2003prejudice}, the addition of identity signals in profiles did not lead to increased levels of received offensiveness, even for identity categories known to be prone to targeted offensiveness such as sexual and gender minorities. Overall, our findings suggest profile-based identity disclosure is an active process signaling future behavior changes in the priorities of a user.

\section{Social Identities and Self-disclosure}
Prior work has examined identity disclosure from the perspectives of language, networks, and social interactions, particularly in online spaces. We build on the previous studies to formulate hypotheses that examine whether disclosure of social identities leads to changes in behaviors of both the user themself and how they are perceived by others.

\begin{figure*}[t!]
    \centering
    \includegraphics[width=\textwidth]{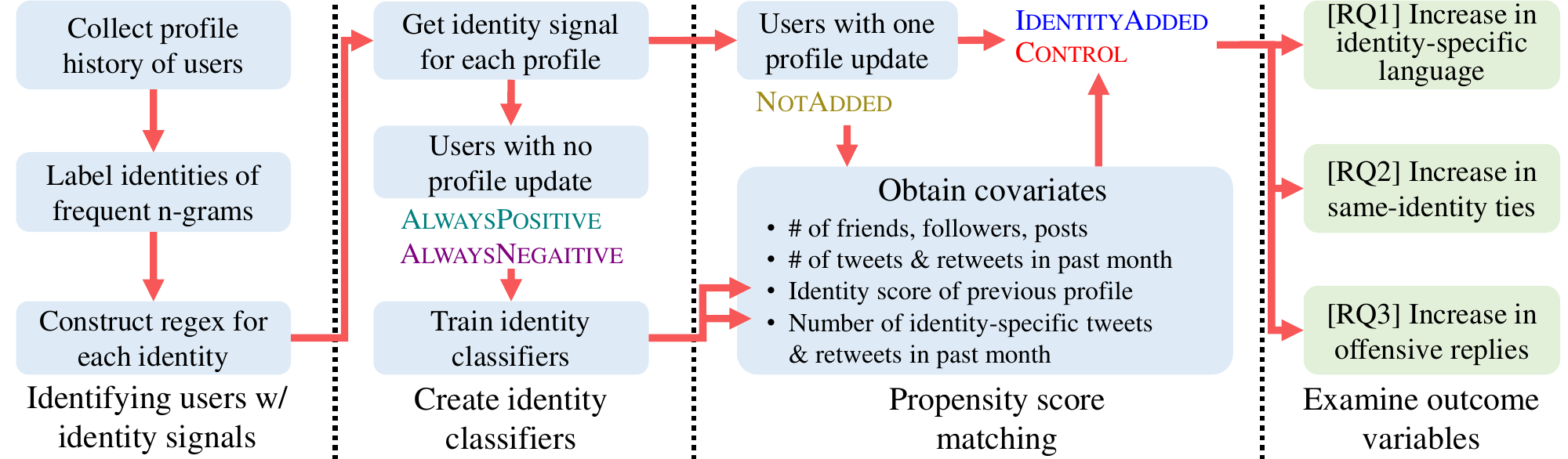}
    \caption{\small Schematic diagram of data collection, identity classifiers, propensity score matching, and analyses on research questions. \alwayspositive, \alwaysnegative, \treated, \untreated and \control represent sets of users with specific profile and identity statuses.}
    \label{fig:flow}
\end{figure*}

\subsection{Social Identities and Language}
Sociolinguistics has long associated language with social identities of the speaker~\citep{labov1966social,eckert2000language,pomerantz2007language}. Specifically, \citet{bucholtz2005identity} propose a framework for understanding identity through linguistic interaction, where they suggest that identities can be indexed through linguistic aspects such as style, stances, and labels~\citep{schilling2004constructing}. This framework also posits that the display of identity through language can be an intentional form of agency to meet social goals~\citep{duranti2008companion}. From this perspective, we can assert that the intention to disclose one's social identity can be reflected through their language, which may be indicative of the identity.


Our first hypothesis examines the relationship between identity disclosed through language and through profile updates. We hypothesize that the modification of one's profile to disclose a particular social identity will motivate the user to tune their linguistic style to accommodate their presented  identity.
\\
\noindent\textbf{H1} \textit{The addition of a social identity on a Twitter profile will  lead to posting more identity-aligned tweets compared to a reference group.}


\subsection{Networked Effects of Identity Disclosure}
People present themselves to others by controlling the amount of information available to maintain a publicly desirable image, a concept known as impression management~\citep{goffman1959presentation}. This management helps achieve socially desirable goals such as maintaining reputation~\citep{schlenker1999impression,zivnuska2005impression}. In social networking platforms such as Twitter or Instagram, the downstream effects of impression management can be translated into measurable outcomes such as maintaining connections with ``friends'' in the platform who can provide desirable effects such as social support or access to information~\citep{lampe2007familiar,yan2022friendcount}. We thus expect that the addition of social identity in one's profile reflects a desire to connect with like-minded others, which results in an increased effort to forge connections with people of the same identity.
\\
\noindent\textbf{H2} \textit{The addition of a social identity on a Twitter profile will directly lead to establishing more network connections with users of the same identity  compared to a reference group.}
\\


\subsection{Consequences of Identity Disclosure}
Identity disclosure can lead to undesirable consequences. Privacy is a major risk of disclosure in online spaces~\citep{ampong2018selfdisclosure}. Also, the disclosure of minority or marginalized identities can lead to being targeted for online harassment. For example, nonbinary users consider disclosing their identity on social media a stressful event~\citep{haimson2015disclosure,haimson2020coming}.

As our final hypothesis, we test whether disclosure of one's identity can lead to increased hostility directed at the user. Specifically, we measure if a user becomes a target of offensive content following the addition of their identity on the profile.
\\
\noindent\textbf{H3} \textit{The addition of a social identity on a Twitter profile will result in receiving more offensive replies  compared to a reference group.}
\\


\section{Identifying Identity Change in Profiles}
\label{sec:data}

Here, we describe our pipeline for identifying instances of Twitter users disclosing social identities on their profiles. An overview of the data collection and processing is shown in Figure~\ref{fig:flow}.

\subsection{Identifying Twitter Profile Changes}
We first identify a set of users who have added signals of their social identity to their Twitter profiles. This information is unobtainable using just the Twitter API as it only returns a user's profile information at the time of the API call and does not provide a chronological timeline of profile changes. We instead use the Twitter Decahose dataset which contains a 10\% sample of the entire Twitter activities over a period of over 12 months. We identify all activities of every user between April 2020 and April 2021.
Each tweet or retweet object includes various metadata, one of which is the user's profile description at the time of the tweet. We collect all instances of user profiles for our Twitter users and sort them in chronological order, enabling us to identify \textit{when} a user changed their profile. We remove verified accounts and users whose language is set to a language other than English, resulting in 15,215,776 users and 73,048,466 unique profiles.

\subsection{Categorizing Social Identities}
Deciding what counts as a social identity can be challenging. 
Here, we start from an initial set of social categories based on two relevant studies.
\citet{priante2016whoami} categorized social identities into five groups: personal relationships, vocations/avocations, political affiliations, ethnic/religious groups, and stigmatized groups. Meanwhile, \citet{yoder2020presentation} used 11 identity categories: age, ethnicity/nationality, fandoms, gender, interests, location, personality type, pronouns, relationship status, sexual orientation, and zodiac. Using this list of categories as a starting point, one author manually inspected each n-gram and assigned it to a category when applicable. The n-grams within each category were additionally grouped into subcategory levels, e.g., the \textit{gender} category consists of three subcategories: \textit{men}, \textit{women}, and \textit{nonbinary}. A total of 221 n-grams were assigned to a category and subcategory. 

We also create subcategory-level identities within each category, which is the basic unit of \textit{social identity} in this study. This process results in a total of ten categories (Table~\ref{tab:identity_counts}) and 44 subcategories of identities (Appendix Table~\ref{tab:identity_counts_subcategory}).
Further details on the subcategories are in Appendix Section~\ref{sec:subcategory}.

After categorizing n-grams into identity categories and subcategories, we follow the approach from prior work~\citep{yoder2020presentation,pathak2021method} and construct regular expressions for each category and subcategory based on the n-grams to improve precision. For example, when constructing regular expressions for age, we ensure that the corresponding phrases include identifiers such as `years old' or `y/o'. 




\begin{table}[t!]
\centering
\small
{
\begin{tabular}{lr}
\hline
\textbf{Category} & \textbf{\# users}  \\
\hline
Age & 12,737 \\
Education & 23,201 \\
Ethnicity & 16,507 \\
Gender pronouns & 59,893 \\
Occupation & 68,694 \\
Personal & 41,107 \\
Political & 26,609 \\
Relationship & 20,167 \\
Religion & 11,169 \\
Sexuality (LGBTQ+) & 3,772 \\
\hline
\textbf{Total} & 283,793 \\ 
\hline
\end{tabular}
}
\caption{\small The number of users who updated their Twitter profiles to disclose social identities. Refer to Table~\ref{tab:identity_counts_subcategory} in the Appendix for counts at subcategory level.}
\label{tab:identity_counts}
\end{table}

Next, we identify a set of users who have changed their profiles to disclose their social identity. We run our regular expressions on every unique profile to determine whether a profile is associated with a particular identity. We assign multiple labels if a user's profile is associated with multiple identity categories (e.g. ``18yo | he/him | father of two wonderful children''), but leave out profiles that our method labels as belonging to multiple subcategories within the same category when they are meant to be mutually exclusive (e.g. age - ``18yo | 30y/o'', political affiliation - ``devout democrat | conservative''). Based on the mapped identities per profile, we can identify all users who satisfy the following two conditions:
(1) each user has made only one change in their profile during the 1 year observation period, and (2) the only change is the addition of a new social identity---i.e., the phrase indicating identity should only exist in the changed profile and not the previous version. This filtering results in a set of 283,793 users who added a single new social identity through Twitter profiles, which we refer to as \treated. Tables~\ref{tab:identity_counts} and \ref{tab:identity_counts_subcategory} contain category- and subcategory-level counts.

We validate the quality of our pipeline for capturing instances of identity disclosure through an annotation task. For each subcategory, three annotators are provided twenty samples which each consist of two subsequent profiles, one pre- and one post-change. The twenty samples include ten positive samples from \treated as well as ten negative samples, which vary from (1) no disclosure in either, (2) disclosure in both, and (3) disclosure only in pre-change. The resulting Krippendorff's $\alpha$ was 0.74, indicating a high level of agreement that the changes detected by our approach do constitute meaningful self-disclosure of identity. We then evaluate our pipeline by evaluating it on the majority vote from the annotations, from which we saw that 41/44 identities achieved an F1 score higher than 0.5 (Appendix Tables ~\ref{tab:pipeline_performance} and \ref{tab:pipeline_performance_all}). We therefore removed the three identities with low performance: education:student, ethnicity:korean, and occupation:art.

\subsection{Propensity Score Matching}
Since our research questions center around behavioral changes following social identity disclosure through profiles, a meaningful measurement can be made by comparing against a control group that displays similar behaviors but does not disclose social identities through profile updates. We adopt propensity score matching (PSM), a quasi-experimental method widely adopted in observational studies involving observational social media data~\citep{yuan2023papageno,choi2023shocks}. 

Apart from the \treated users we also identify 849,901 users who (1) made one profile update during the 1-year observation period but (2) did not include any phrases of social identity in their profiles before or after the update, which we refer to as \untreated users. For each user in \treated and \untreated, we identify the following covariates obtained at the date of the profile change: number of days since account creation, number of friends, number of followers, number of total posts, number of tweets and retweets posted during one month prior to the time of profile update. Further details of the matching can be found in Appendix Section~\ref{sec:sub:matching}.

As a result of the matching process, we are left with 283,566 treated users and 1,228,945 matched users.  We refer to the resulting matched set as \control users. Figure~\ref{fig:smd} in the Appendix shows that the distribution containing the standardized mean difference of every covariate reduces sharply after matching, demonstrating the diminished effect caused by confounding covariates.

\subsection{Estimating Treatment Effects}
\label{sec:sub:equation}
Our setting of treated and control variables allows us to perform a widely used causal inference method known as difference-in-differences ~\citep[DiD; ][]{abadie2005diffindiff}. 
Though DiD is most commonly used when the outcome variable is a continuous variable, it can be applied to different types of outcomes such as count variables~\citep{cameron2013regression,mark2013california}. Accordingly, we use the following equation:

\begin{math}
    \textrm{log}\left( y_{i,t} \right)=\beta_0+\beta_1X_{i}+\beta_2\left( T=1 \right)+\beta_3\left( t\ge \textrm{tr} \right) +\beta_4\left( T=1 \right)\left( t\ge \textrm{tr} \right)
\end{math}

\noindent
where $y_{i,t}$ is the outcome variable at time $t$ for user $i$, $T=1$ is a binary assignment status to treatment group, and $t>=\textrm{tr}$ is whether time $t$ is beyond treatment period. $X_i$ is the time-invariant covariates of $i$, which consist of the number of friends, followers, and total posts.
All experiments are modeled as a negative binomial regression using generalized estimating equations (GEE) in \texttt{statsmodels}. Because our hypothesis testing are done on multiple identities, we apply the Bonferroni-Holm correction~\citep{holm1979simple} to account for false positives when reporting significance test results from the regressions.




\section{How does Identity Disclosure affect Language?}
To understanding behavioral changes following identity disclosure, we first study whether users change their language following profile updates to include a social identity. We hypothesize that the addition of an identity signal provides a certain level of boost to represent their identity more through the content they produce and engage with.

\paragraph{Measuring Identity-specific Language}
We first construct classifiers to measure the amount of identity alignment from a tweet.
Based on existing findings that posts and profile descriptions in online platforms are reflective of one's social identity~\citep[e.g.,][]{priante2016whoami,preotiuc2018race}, we assume that if a user has disclosed a social identity on their profile description for a sufficiently long time, then the text created by the user contains topical and stylistic features indicative of the disclosed identity. Accordingly, we first identify users who \textit{did not} update their profile during our observation period, and identify cases where their profile did~(\alwayspositive) or did not~(\alwaysnegative) include an identity~(refer to Figure~\ref{fig:flow}). We then aggregate the tweets created by each user and assign positive or negative labels to the tweets based on the user's identity existence. Each classifier is a RoBERTa~\citep{liu2019roberta} model pretrained from tweets and further finetuned on the labeled tweet dataset. Further training details and examinations on classifier performances can be found in Appendix Section~\ref{sec:sub:classifier}.

\paragraph{Experiment Setting}
We use the scores from the classifiers to measure levels of identity-specific language from both the content that users post (tweets) and engage with through sharing (retweets).
Using the identity classifiers, we obtain scores for every tweet and retweet generated by each \treated and \control user between one month before and after the profile update. We then count the number of tweets with an inferred identity score higher than 0.5 and aggregate them into two periods, before and after the profile update. We consider these as the total number of identity-relevant tweets the user tweeted or retweeted before or after treatment. We also count the number of total tweets regardless of identity score, which captures overall activity levels. We run separate regressions with the number of total tweets/retweets and identity-specific tweets/retweets as outcome variables, and include the number of total activities as a control variable when modeling identity-specific activities.

\begin{figure*}[t!]
    \centering
    \includegraphics[width=\textwidth]{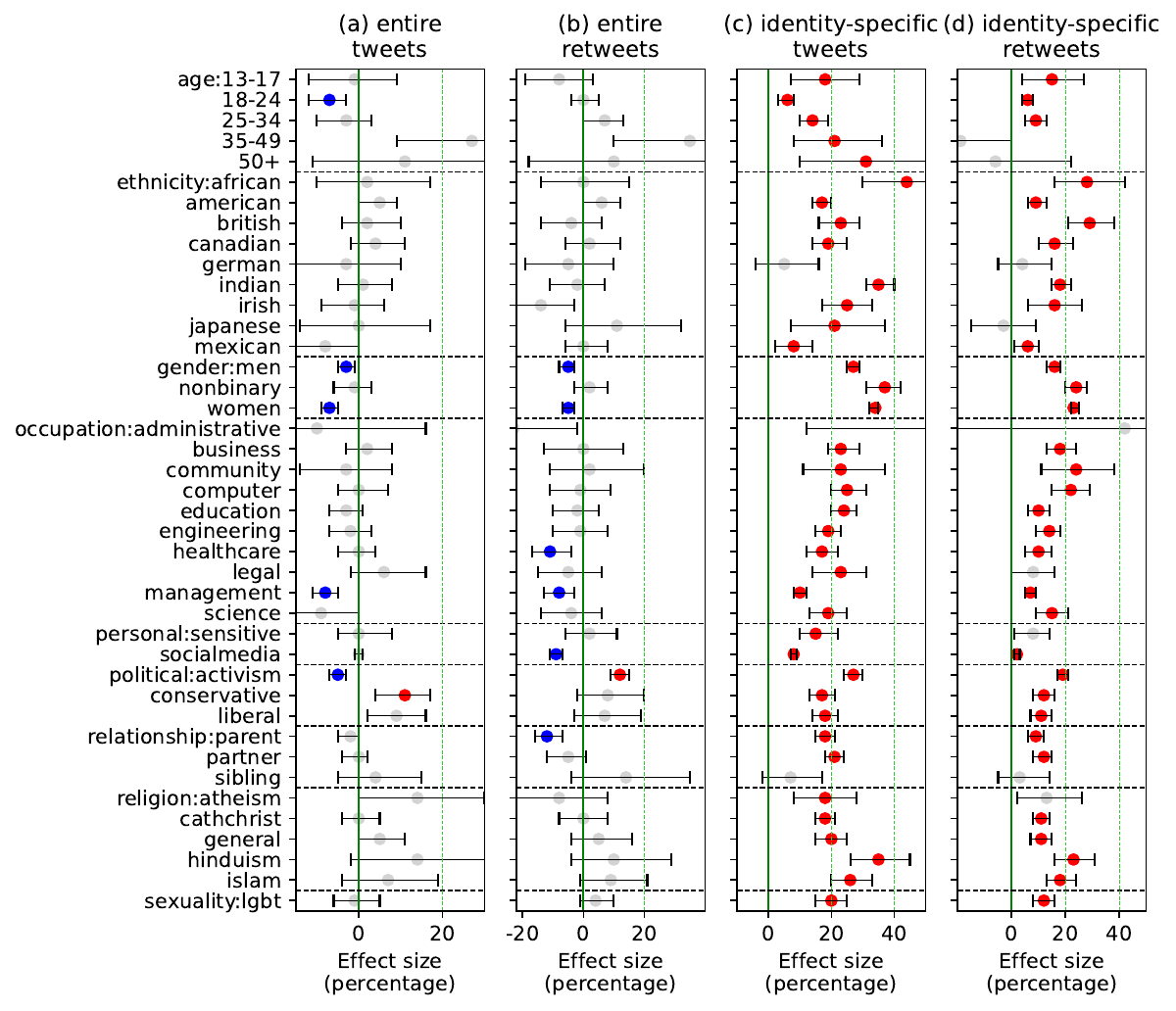}
    \caption{\small Effect sizes of identity disclosure on tweet and retweet-level activities. The x-axis indicates percentage increase in number of tweets following identity disclosure. Significant positive and negative values that pass the correction test are marked in red (positive) and blue (negative). While identity disclosure does not lead to increased activity levels, there are significant increases in the number of tweets and retweets that contain identity-specific language}
    \label{fig:identity_tweet_effect}
\end{figure*}

\paragraph{Results}
Figure~\ref{fig:identity_tweet_effect} shows the effects of adding profiles on four different types of tweet activity counts: the number of total tweets (Figure~\ref{fig:identity_tweet_effect}(a)) and retweets (Figure~\ref{fig:identity_tweet_effect}(b)) versus identity-aligning tweets (Figure~\ref{fig:identity_tweet_effect}(c)) and retweets (Figure~\ref{fig:identity_tweet_effect}(d)). We can first observe that, 
contrary to prior work~\cite{lampe2007familiar}, the additional disclosure of social identity via profiles does not lead to greater overall activity levels compared to profile updates without such disclosure (Figures \ref{fig:identity_tweet_effect}(a) and \ref{fig:identity_tweet_effect}(b)). In fact, we observe the opposite for several types of identities, most notably drops of both tweet and retweet levels in binary gender pronouns and student status. The only statistically significant increases we observe arise from disclosing political statuses.

On the other hand, we observe statistically significant increases in the number of tweets posted and retweeted which contain identity-specific language, across \textit{almost every} category~(Figures \ref{fig:identity_tweet_effect}(c) and \ref{fig:identity_tweet_effect}(d)). Though there exists variance among categories, in general, we observe that identity-specific tweets increased by around 20-40\% and identity-specific retweets increased by around 10-30\%, indicating that though the content volume does not change, the percent of identity-related content within that volume increases substantially.
Further comparisons within identity categories reveal interesting findings. For instance, we observe that for both tweets and retweets, the increase following identity disclosure of men is lower than that of women and nonbinary genders. One possible reason is women and nonbinary gender users may undergo harder decisions to disclose their identity, which results in a greater change in their behavior following disclosure. Similarly, our results on ethnicity disclosures show larger identity-specific activities for African identities compared to the American identity, suggesting the level of language change may differ by identity types.

\begin{figure}[t!]
    \centering
    \includegraphics[width=0.7\textwidth]{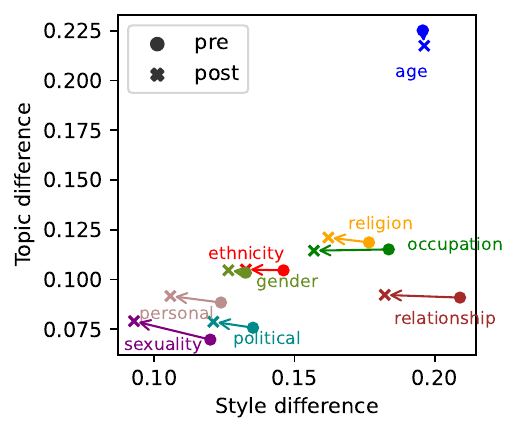}
    \caption{\small Changes in style and topic differences between \alwayspositive users and \treated users before and after identity disclosure. Style becomes more similar after the disclosure compared to topics, where the relative distances are much smaller.}
    \label{fig:lang_change}
\end{figure}

\paragraph{Identity-specific language: topic or style?}
To further understand which aspects of language change following identity disclosure, we compare the tweets through two components of language: \textit{topic} and \textit{style}. We examine whether having a \treated user disclose their identity results in their language becoming more similar to that of a \alwayspositive user regarding each component. Further details for computing the distances can be found in Appendix Section~\ref{sec:sub:style}.

Figure~\ref{fig:lang_change} shows changes in the distance between the language of users who change towards disclosing their identity to those who always have had the identity visible. While topic differences remain relatively unchanged, the difference in style between the two user groups are reduced following identity disclosure for all categories apart from age. Though users do not significantly shift their topics of interest, they tune their language to appear more similar to the style associated with the identity that they choose to disclose.

\section{Does identity disclosure in profiles lead to network rewiring towards same-identity connections?}

\begin{figure}[t!]
    \centering
    \includegraphics[width=0.8\textwidth]{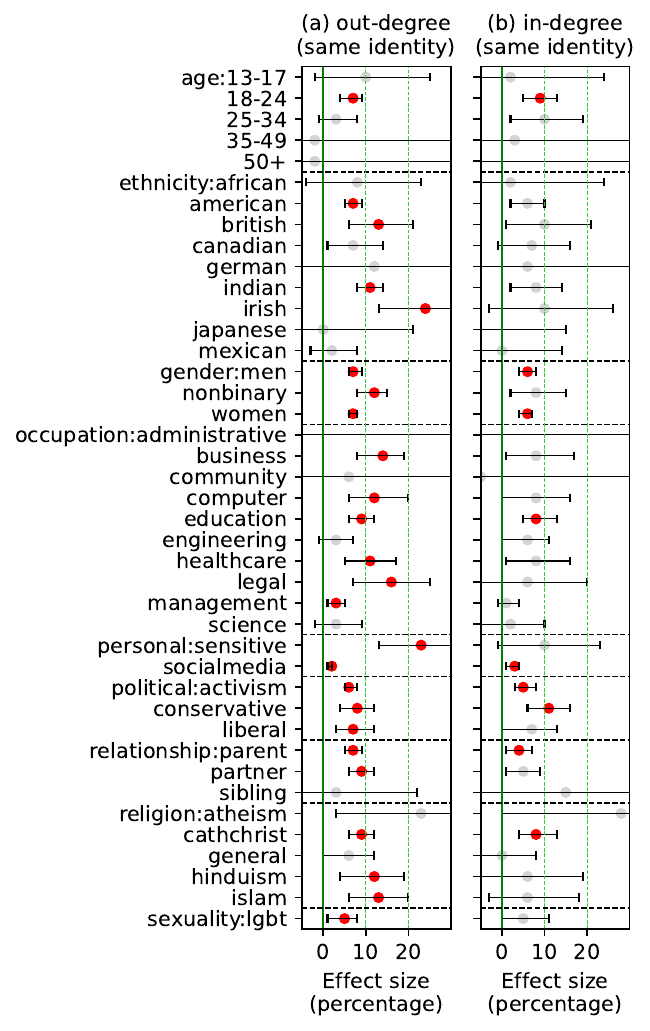}
    \caption{\small Effect sizes of identity disclosure on out- and in-degree network sizes. Users reach out to those of the same identity following disclosure (out-degree), but not all identities receive increased attention from others in return (in-degree)}
    \label{fig:diff_network}
\end{figure}

In our next analysis, we investigate whether the addition of identities leads to bridging more connections with like-minded others. To do so, we collect the ego networks of every \treated and \control user where an edge between two users $u$ and $v$ is defined when $u$ replies to or retweets a tweet posted by $v$. We divide a user's network activities by pre- and post-treatment where we look at a timespan of 12 weeks. We use the same set of regular expressions from the profiles of all users included in the networks and extract any social identities from their profiles during the 12-week period. The subset of connected users who have adopted the same identity as the ego user at any point will be considered same-identity nodes. Thus, in our subsequent diff-in-diff analysis, the outcome variable is the number of same-identity nodes before and after the identity disclosure.

\paragraph{Results}
Figure~\ref{fig:diff_network} displays the treatment effect on the out- and in-degree of the network when restricted to users of the same identity. We can observe that across most categories, the out-degree of same-identity neighbors significantly increases after identity disclosure in profiles~(Figure~\ref{fig:diff_network}(a)). This indicates that the users who choose to disclose their identities also choose to connect to more people that share the same identity.

We next look at the in-degree level changes, which is a stronger indicator of how the addition of identity is viewed by others~(Figure \ref{fig:diff_network}(b)). We observe that the in-degree of same-identity groups is less likely to increase compared to the out-degree, which indicates that inbound connections are less likely to be made compared to outbound connections, as the former requires others to actually be motivated to establish new connections with the user who has made a profile change. 

Additional results, shown  in the Appendix, highlight identity-specific changes. Figure~\ref{fig:net_all} contains the effect sizes of the \textit{total} out- and in-degree network sizes following disclosure, revealing that the overall network size only increases for political identities. These results support our claim that users choose to strategically rewire their connections more towards those of the same identity while keeping overall network sizes stable instead of merely being more open in general. Figure~\ref{fig:net_other} shows changes in connection levels towards different identities in the \textit{same} category. We find that gender pronouns is the only category to increase in both in-degree and out-degree for all identities, which is in line with existing work that showed tie clustering among such pronouns~\citep{tucker2023pronoun}. Last of all, we compare changes in cross-partisan connections for conservative and liberal users, where we observe significant increases of outbound connections from those who disclose their liberal identity to conservative users, but not the other way round.


\section{Does identity disclosure lead to receiving more offensive content?}
In our final research question, we investigate possible negative consequences of disclosing one's identity, namely whether identity disclosure leads to increased targeted offensive content.

\paragraph{Experiment Setting}
For each \treated and \control user, we use the 10\% sample dataset to collect a history of the tweets posted by the user during one month before and after the time of their profile update, as well as all replies received from other users during this period. Next, we use a publicly available classifier for detecting offensiveness from Hugging Face~\citep{barbieri2020tweeteval}{\footnote{https://huggingface.co/cardiffnlp/twitter-roberta-base-offensive}} to obtain offensiveness scores of both the tweets posted and the replies from others. We then formulate an equation to model the expected number of offensive replies

\begin{math}
    \textrm{log}\left( y_{i,t} \right)=\beta_0+\beta_1X_{i}+\beta_2\left( T=1 \right)+\beta_3\left( t\ge \textrm{tr} \right) +\beta_4\left( T=1 \right)\left( t\ge \textrm{tr} \right) + \color{blue}\textrm{log}\left( \beta_5\textrm{n}_{id} \right)  +  \color{red}\textrm{log}\left( \beta_6\textrm{n}_{id} \right) \left( T=1 \right) \left( t\ge \textrm{tr} \right)
\end{math}. \\
The added term $\textrm{log}\left( \beta_5\textrm{n}_{id} \right)$ indicates the log-normalized number of identity-specific tweets posted by the user and $\textrm{log}\left( \beta_6\textrm{n}_{id} \right) \left( T=1 \right) \left( t\ge \textrm{tr} \right)$ is the interaction effect between identity disclosure via profile and identity-specific tweets.

\begin{figure}[t!]
    \centering
    \includegraphics[width=0.8 \textwidth]{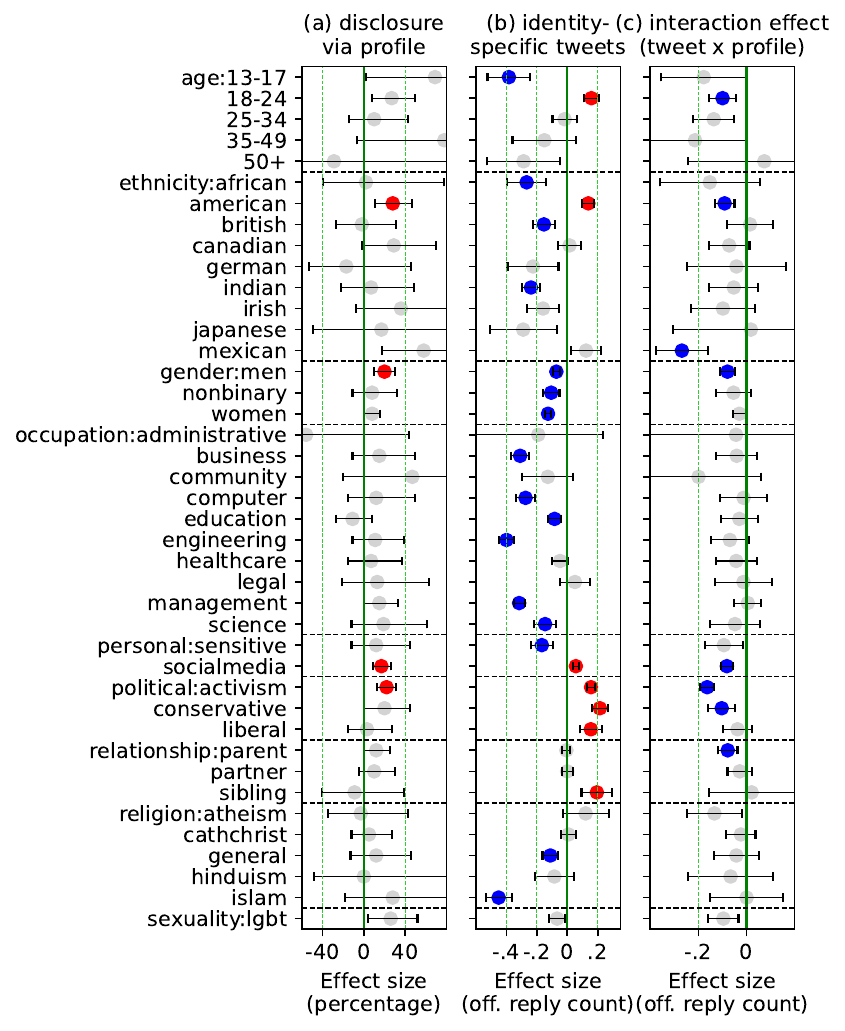}
    \caption{\small Effect size of identity disclosure on the number of offensive replies received. (left) identity addition to profile, (middle) number of identity-specific tweets per week, (right) interaction effect of identity disclosure through profile and number of identity-specific tweets per week}
    \label{fig:diff_offensive}
\end{figure}

\paragraph{Results}
Figure~\ref{fig:diff_offensive}(a) ($\beta_4$) first shows that identity disclosure through profiles increases offensiveness for only a handful of categories - ethnicity:American, gender:men, personal:socialmedia, and political:activism. However, when we observe changes in offensiveness levels caused by increased identity of tweets (Figure~\ref{fig:diff_offensive}(b) ($\beta_5$)), we can see that significant effects can be seen from several categories. Interestingly, the disclosure of identity through tweets leads to reduced levels of offensiveness from others for the three studied gender types, as well as for occupations and religion types. Meanwhile, we observe increased levels of offensive replies from all three types within the political category, hinting that this may be due to heated political conversations that often correlate with offensiveness. Lastly, the interaction effect of identity disclosure via both tweet and profile (Figure~\ref{fig:diff_offensive}(c)) ($\beta_6$) suggests that the combined effect from disclosure through both channels reduces levels of offensiveness for every category where increased identity disclosure through tweets was associated with increased offensiveness. One potential explanation is that disclosing identity through both profile and tweet could create a sense of consistency, which helps reduce levels of hostility towards that identity group.


\section{Discussion}
Our findings indicate that disclosing social identities, regardless of category, follows similar behaviors in that both the content produced and connections made by the user become more aligned with the announced identity. We can assume that at the heart of such disclosure lies the innate desire to express oneself and find comfort among like-minded peers. It is also notable that instead of just becoming more active overall, users maintain similar levels of activity and connectivity while \textit{channeling} their effort towards more identity-aligned decisions. This comes at the expense of interactions with those unassociated with the identity, and coupled with existing findings that more people are disclosing their identity on Twitter~\citep{pathak2021method,jones2021twitterbio}, could even signal that our Twitter networks might become more homogeneous over time.

Another interesting finding regarding the effects of disclosure was that identity disclosure via profiles did not result in significant increases in offensive replies targeted to the user for marginalized categories such as nonbinary gender, LGBTQ+ sexualities or minority ethnicities. While our results do not and are not meant to deny the existence of identity-targeted hate in social media platforms that is a major source of harm, we take a more positive view suggesting that the consequences of disclosing identity through profiles may not be as severe as anticipated, and that disclosure should be promoted and more widely accepted.



\section{Conclusion}
We conduct a case study for identifying whether added disclosure of one's social identity through profile updates leads to subsequent changes in linguistic style and network connections, and whether the disclosure leads to increased offensiveness from others. We propose methods for measuring identity disclosure through both profile- and tweet-level language and apply them to quasi-experimental difference-in-differences methods to show that identity disclosure through labels leads to increased disclosure in subsequent language. Furthermore, we also observe that identity disclosure can lead to increased connections with like-minded identities, which is much more prevalent from the outward versus inward ties. Finally, we observe that, contrary to existing concerns, the negative effect of increased offensiveness from disclosing a social identity via profiles does not exist for most identities, and that the combined disclosure from both profile- and tweet-levels led to reduced targeted offensiveness levels. Overall, our results suggest that the decision to disclose one's social identity can be encouraged, with negative effects appearing less than is concerned. The code and annotated data for the study will be available at \textit{https://github.com/minjechoi/twitter\_identity}.




\section{Limitations and Ethical Considerations}
One limitation that our analysis is focuses only on Twitter. The amount of disclosure may differ by type of platform depending on why people use it~\citep{jaidka2018facebook,vandijck2013facebook}.
Identity may be visible through means other than the profile text. One example would be a profile image, which can indicate demographic features such as age, gender, and ethnicity~\citep{yoder2020presentation}.

\paragraph{Identity-unaware offensiveness classifiers}
To conduct the experiment on offensiveness levels after identity disclosure, we use finetuned classifiers trained on an external Twitter corpus~\citep{barbieri2020tweeteval}. The black-box nature of these classifiers and datasets contain the risk of predicting text features of some identities as more offensive than others without sufficient understanding of contexts surrounding the identity, such as African-American English~\citep{sap2019risk,harris2022aae}. In fact, our correlation results between the scores of the offensive classifier and identity-specific classifiers on a large corpus (Figure~\ref{fig:style} in the Appendix) may lead to conclusions such as identity-specific language from nonbinary genders being more likely to be offensive than men or women, or the identity-specific language of Mexicans being the most offensive compared to other ethnicities.

\paragraph{Purpose of identity classifier}
It is possible that one may associate the regular expression-based pipelines for identifying profile disclosures and the identity classifier models with purposes such as detecting whether a user possesses a hidden identity trait based on their prior Twitter history. We argue that our models are not served for that purpose. Rather, our categorization of users is entirely based on self-declared phrases indicative of social identities, which we examine through a meticulous verification process. Our results are driven from purely observational data aggregated at a scale of hundreds or thousands of users, which removes the possibility of identification.

\paragraph{Results on the disclosure of marginalized identities}
One of the findings of our study is that the disclosure of social identities via profile changes did not result in increased levels of targeted offensiveness, even for marginalized identity groups such as specific gender or ethnicity groups. One possible limitation is that our study is based on users who have willingly made the decision at some point to update their profile and make their identity visible to their friends and to the public, and those who did update may have been in a situation where they felt more comfortable to disclose in the first place. This creates a selection bias that might interfere with the generalizability of our findings to the general population of Twitter users, and thus further caution should be made when estimating the reactions following disclosure in online spaces. Nevertheless, we conclude from our findings that identity disclosure through profiles can be an effective means of expressing oneself and connecting with like-minded others, and would encourage users to do so if seeking such outcomes.

\bibliography{main}
\bibliographystyle{acl_natbib}

\clearpage

\appendix

\begin{table}[]
\centering
 \resizebox{0.42\textwidth}{!}{
\begin{tabular}{llr}
\hline
\textbf{Category} & \textbf{Subcategory} & \textbf{\# users} \\
\hline
Age & 13-17 & 871 \\
    & 18-24 & 8,872 \\
    & 25-34 & 2,449 \\
    & 35-49 & 381 \\
    & 50+   & 164 \\
\hline
Education & Student & 23,201 \\
\hline
Ethnicity & African & 575 \\
          & American   & 5,397 \\
          & British & 1,487 \\
          & Canadian & 2,050 \\
          & German & 636 \\
          & Indian & 3,045 \\
          & Irish & 1,023 \\
          & Japanese & 349 \\
          & Korean & 259 \\
          & Mexican & 1,686 \\
\hline
Gender pronouns & Men & 19,115 \\
       & Women & 36,708 \\
       & Non-binary & 4,070 \\
\hline
Occupation & Administrative & 160 \\
           & Art & 28,746 \\
           & Business & 3,284 \\
           & Community & 635 \\
           & Computer & 3,031 \\
           & Education & 6,556 \\
           & Engineering & 4,765 \\
           & Healthcare & 4,109 \\
           & Legal & 1,117 \\
           & Management & 13,646 \\
           & Science & 2,645 \\
\hline
Personal & Social Media & 39,310 \\
         & Sensitive & 1,797 \\
\hline
Political & Conservative & 2,059 \\
          & Liberal & 2,347 \\
          & Activism & 22,203 \\
\hline
Relationship & Partner & 6,966 \\
         & Parent & 12,233 \\
         & Sibling & 968 \\
\hline
Religion & Catholic / Christian & 5,954 \\
         & Islam & 1,255 \\
         & Hinduism & 544 \\
         & Atheism & 387 \\
         & General & 3,029 \\
\hline
Sexuality & LGBTQ+ & 3,772 \\
\hline
\textbf{Total} & & 283,793 \\ 
\hline
\end{tabular}
}
\caption{Count of users who added social identities to their Twitter profiles once in our observation period for each subcategory-level identity.
}
\label{tab:identity_counts_subcategory}
\end{table}

\section{Details on Category- and Subcategory-level Identity Categorization}
\label{sec:subcategory}
We start from an initial set of social categories based on two relevant studies.
\citet{priante2016whoami} categorize social identities into five groups based on the findings of \citet{deaux1995parameters}: personal relationships, vocations/avocations, political affiliations, ethnic/religious groups, and stigmatized groups. Meanwhile, \citet{yoder2020presentation} constructed identity categories based on \citet{bucholtz2005identity}: age, ethnicity/nationality, fandoms, gender, interests, location, personality type, pronouns, relationship status, sexual orientation, and zodiac. Using this list of categories as a starting point, one of the authors manually inspected each n-gram and assigned it to a category when applicable. Furthermore, the n-grams within each category were additionally grouped into subcategory levels. For instance, the \textit{gender} category consists of three subcategories: \textit{men}, \textit{women}, and \textit{nonbinary}. A total of 221 n-grams were assigned to a category and subcategory. 
A list of the categories and subcategories can be found in Table~\ref{tab:identity_counts_subcategory}. Descriptions of the final categories are as follows:
\begin{itemize}
    \item \textbf{Age} This category contains the disclosed age of the user. We grouped age into five bins to represent teenagers (13-17), college students (18-24), young adults at early stages of their careers (24-35), adults at the age of parenthood and advanced careers (35-49), and senior adults (50+). We aknowledge different categorizations of age could be used in this study, such as that of \citet{wang2019m3}.
    \item \textbf{Education} We constructed a single-identity category \textit{Education} to collect instances of students disclosing their education status, such as degree name, current university, or school year.
    \item \textbf{Ethnicity} This category contains the self-declared ethnicity of the user. We included words or phrases describing the user's ethnicity as well as nationality flag emojis which can be used to describe one's nationality. Our subcategories are limited to countries where there was at least one corresponding n-gram.
    \item \textbf{Gender pronouns} Following the work of \citet{jiang2022pronouns}, we use three subcategories of gender pronouns: men, women, and nonbinary.
    \item \textbf{Occupation} Occupation categories were obtained from the International Standard Classifications of Occupations (ISCO-08) list, where we selected all sub-major group categories which corresponded to any of the top n-grams we examined.
    \item \textbf{Political} This category corresponds to the disclosed political leaning of the user. Along with subcategories for conservative and liberal, we include another category related to activism, which in this case corresponds to phrases related to the Black Lives Matter movement.
    \item \textbf{Relationship} Based on the frequent n-grams, we identify three types of family relationship types mentioned in profiles: \textit{partner}, \textit{parent}, and \textit{sibling}.
    \item \textbf{Religion} We identify n-grams containing religious terms, and create subcategories for each different religion that was mentioned. For Christianity and Catholism we discover that it is difficult to split out the two and thus combine them into a single category. Finally, n-grams genuinely mentioning `God' are mapped to the \textit{General} subcategory
    \item \textbf{Sexuality} We identified n-grams corresponding to LGBTQ+ identities and map them into a single subcategory. We remove phrases that signal only indirect membership (e.g. LGBT-ally)
    \item \textbf{Personal} We define a category for two additional types of self-disclosure. One is the disclosure of additional social media accounts, and the other is that of stigmatized identities such as joblessness, health issues, and trauma.
\end{itemize}

\begin{table}[t!]
\centering
{
\begin{tabular}{lr}
\hline
\textbf{Category} & \textbf{F1 score}  \\
\hline
Age & 0.81 \\
Education & 0.46 \\
Ethnicity & 0.78 \\
Gender pronouns & 0.98 \\
Occupation & 0.76 \\
Personal & 0.71 \\
Political & 0.85 \\
Relationship & 0.82 \\
Religion & 0.86 \\
Sexuality (LGBTQ+) & 0.82 \\
\hline
\end{tabular}
}
\caption{Performance on detecting an identity disclosure from profile changes, aggregated at category level. The performance of our pipeline is evaluated against the majority vote of the annotators. We achieve reasonable performance across all categories except for Education, which we remove from subsequent analyses.}
\label{tab:pipeline_performance}
\end{table}

\begin{table}[t!]
\centering
\small
{
\begin{tabular}{llrr}
\hline
\textbf{Category} & \textbf{Identity} & \textbf{Kripp.} & \textbf{F1 score}  \\
\hline
Age & 13-17 & 0.68 & 0.67 \\
Age & 18-24 & 0.71 & 0.82 \\
Age & 25-34 & 0.87 & 0.95 \\
Age & 35-49 & 0.79 & 0.78 \\
Age & 50+ & 0.85 & 0.82 \\
\hline
\color{red}Education & \color{red}student & \color{red}0.17 & \color{red}0.46 \\
\hline
Ethnicity & African & 1.0 & 0.57 \\
Ethnicity & American & 0.57 & 0.75 \\
Ethnicity & British & 0.81 & 1.0 \\
Ethnicity & Canadian & 1.0 & 0.89 \\
Ethnicity & German & 0.53 & 0.89 \\
Ethnicity & Indian & 0.83 & 1.0 \\
Ethnicity & Irish & 0.65 & 0.75 \\
Ethnicity & Japanese & 0.36 & 0.57 \\
\color{red}Ethnicity & \color{red}Korean & \color{red}-0.02 & \color{red}0.0 \\
Ethnicity & Mexican & 0.67 & 0.89 \\
\hline
Gender & men & 0.93 & 1.0 \\
Gender & nonbinary & 0.73 & 0.95 \\
Gender & women & 0.87 & 1.0 \\
\hline
Occupation & administrative & 0.73 & 0.89 \\
\color{red}Occupation & \color{red}art & \color{red}0.41 & \color{red}0.31 \\
Occupation & business & 0.66 & 0.75 \\
Occupation & community & 0.55 & 0.82 \\
Occupation & computer & 1.0 & 0.75 \\
Occupation & education & 0.55 & 0.82 \\
Occupation & engineering & 0.72 & 0.95 \\
Occupation & healthcare & 0.71 & 0.82 \\
Occupation & legal & 0.62 & 0.75 \\
Occupation & management & 0.84 & 0.75 \\
Occupation & science & 0.8 & 0.57 \\
\hline
Personal & sensitive & 0.61 & 0.82 \\
Personal & social media & 0.21 & 0.57 \\
\hline
Political & activism & 0.51 & 0.95 \\
Political & conservative & 0.93 & 0.82 \\
Political & liberal & 0.58 & 0.75 \\
\hline
Relationship & parent & 0.84 & 0.75 \\
Relationship & partner & 0.93 & 0.95 \\
Relationship & sibling & 0.78 & 0.75 \\
\hline
Religion & atheism & 0.8 & 1.0 \\
Religion & Catholic / Christian & 0.45 & 0.67 \\
Religion & general & 0.8 & 0.95 \\
Religion & Hinduism & 0.93 & 0.89 \\
Religion & Islam & 0.85 & 0.75 \\
\hline
Sexuality & LGBTQ+ & 0.78 & 0.82 \\
\hline
\end{tabular}
}
\caption{Performance on detecting an identity disclosure from profile changes for each individual identity. The Krippendorff's alpha agreement score is computed from the results of the three annotators. The performance of our pipeline is evaluated against the majority vote of the annotators. The three identities with low performance are removed from subsequent analyses.}
\label{tab:pipeline_performance_all}
\end{table}

\section{Identifying Identity-specific Language on Twitter}
\label{sec:sub:classifier}
Our analyses require models to quantify language that aligns with a particular social identity. We aim to achieve this by formulating classification tasks to distinguish the language patterns between two types of users. 

\subsection{Experiment setting}
For each identity subcategory, we define \alwayspositive users as those who (1) did not make any changes to their profiles during our observation period, and (2) contained phrases of a specific identity type in their profile. Similarly, we define \alwaysnegative users as those who (1) did not make any changes, and (2) did not include any identity-specific phrases in their profile. Here we assume that the tweets posted by a user with an identity-specific phrase in their profile are more likely to align with the listed identity, and so use the labels of the user as proxies for the tweets. However, it would be unrealistic to assume that all tweets contain such alignment. Therefore, for each user, we assign positive/negative labels at corpus level instead of the individual tweet-level, where each sample consists of a corpus of five randomly sampled tweets posted by a user. We restrict our tweets to those that have user-generated text other than URLs, which we replace with a \texttt{[URL]} token. To distinguish the different texts, a \texttt{</s>} separation token is inserted between each tweet.

For each identity subcategory, we sample up to 50K positive and 50K negative users, which we split into train/test/validation sets on an 8:1:1 ratio with balanced positive/negative samples. For identity classes with insufficient positive samples, we allow each user to be represented in up to ten different samples provided they have enough unique tweets.
We allow for upsampling on identities with small sample sizes on the training set. We finetune each identity separately using a RoBERTa model pretrained on a Twitter corpus~\citep{barbieri2020tweeteval} provided via the Hugging Face API. The training is done on Pytorch 1.13 and Pytorch Lightning 1.8.6 on an NVidia A5000 machine. We use a learning rate of 1e-6 after 100 initial warmup steps followed by linear decay and run for a maximum of 10 epochs where we stop if the validation performance measured in AUC does not increase after two consecutive epochs.

\subsection{Model performances}
\begin{figure}[t!]
    \centering
    \includegraphics[width=0.7 \textwidth]{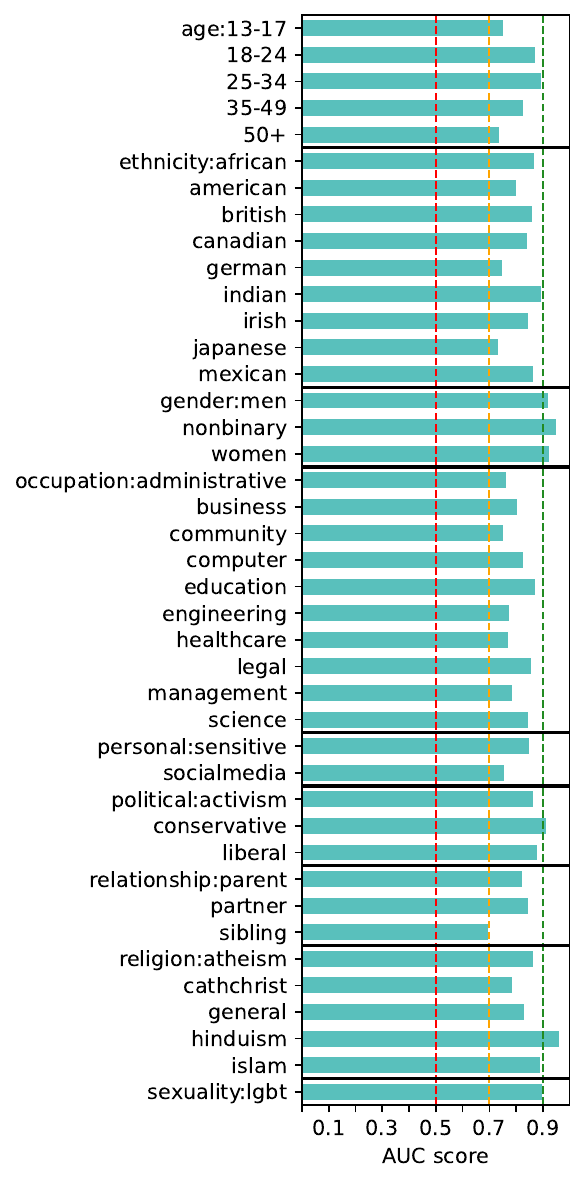}
    \caption{AUC scores of identity-specific language classifiers on test set. Almost all of our categories exceed 0.7, a reasonable cutoff for binary classification.}
    \label{fig:auc}
\end{figure}

\begin{figure}[t]
    \centering
    \includegraphics[width=0.7 \textwidth]{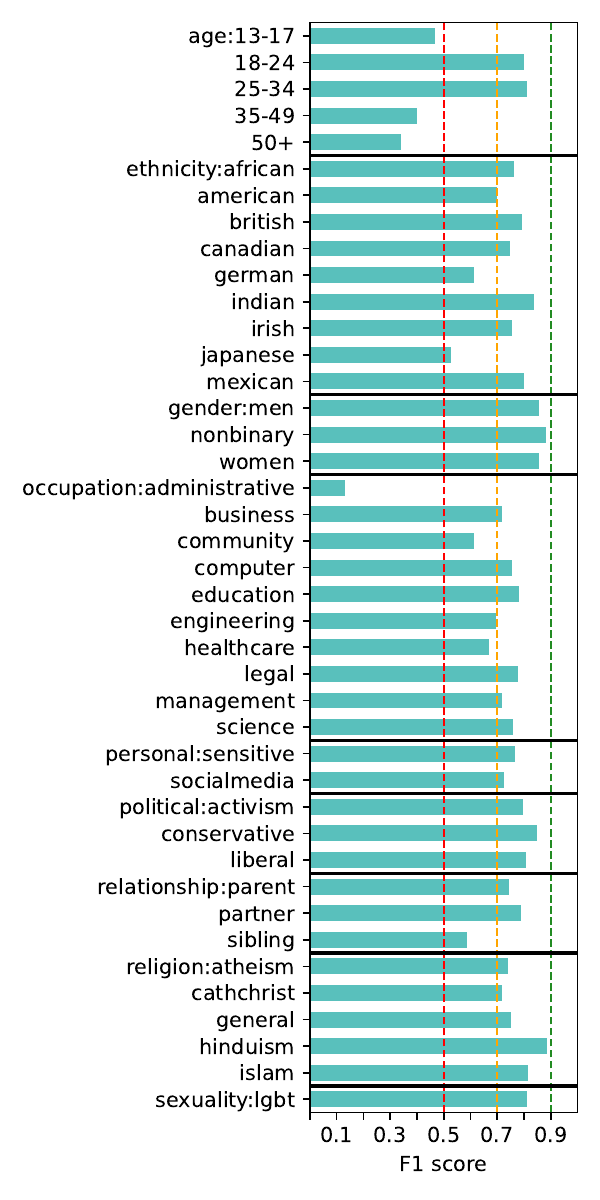}
    \caption{F1 scores of identity-specific language classifiers on test set. While most tasks have a relatively high F1 score above 0.7, some identities are harder to be predicted correctly in a binary setting.}
    \label{fig:f1}
\end{figure}

We evaluate the performances of all models using two metrics: AUC and F1 score. AUC scores are generally high, with all models exceeding a performance of 0.7 (Figure~\ref{fig:auc}). This indicates that the models are doing a good job at assigning higher scores to tweets that contain more signals of identity and vice versa. F1 scores are lower in general, with a few identities such as age:35-49 and occupation:administrative performing worse than random~\ref{fig:f1}. The results from these two figures combined together indicate that while the model sometimes struggles predicting the correct label (positive/negative) for some identities, overall it does a decent job in producing continuous scores which we can use for measuring strong and weak associations of certain identities from texts. Therefore, we proceed with using all of the classifiers for subsequent experiments.

\subsection{Cross-identity similarities in language}
\begin{figure*}[t!]
    \centering
    \includegraphics[width=0.7\textwidth]{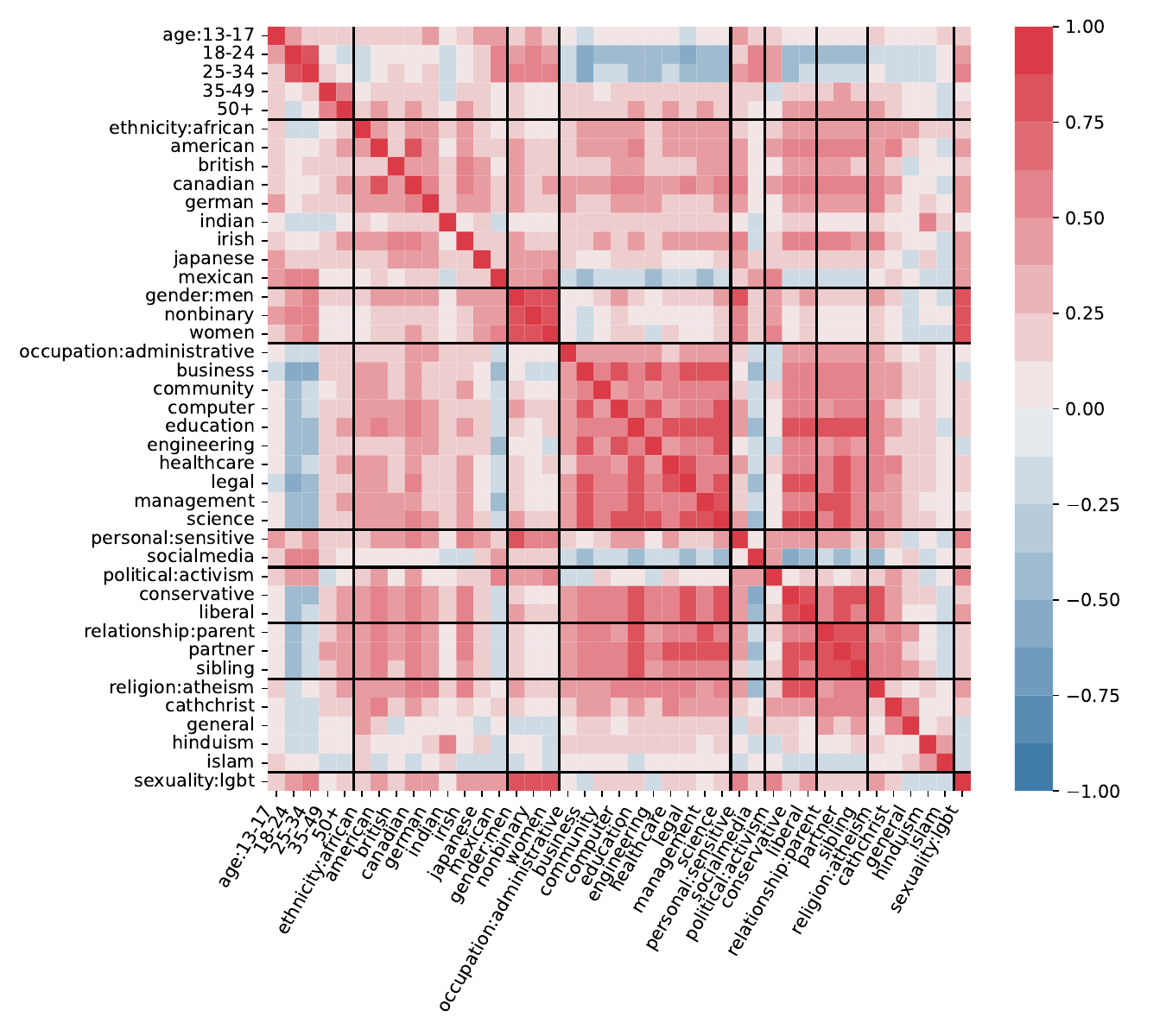}
    \caption{Pairwise comparison of the identity classifiers. Each identity classifier was used to obtain the identity scores from an identical dataset of 1 million randomly sample tweets. Spearman's rank was used to obtain the pairwise similarities between the score distributrions of any two identities. Pairwise similarities are largest between within-category comparisons, indicating that the language associated with identity disclosure follows some categorical properties as well.}
    \label{fig:corr}
\end{figure*}

We compare the pairwise similarities between the identity-specific languages across different identities. We first sample a large corpus of one million random tweets from the history of tweets by \alwaysnegative users, so that we avoid biasing our tweets towards any particular identity. Next, we obtain the identity scores for each identity by running every classifier on the same corpus. We compute Spearman rank pairwise similarity between all identity pairs.
Figure~\ref{fig:corr} contains the pairwise scores for all pairs. We can observe stronger similarity scores for within-category comparisons. This suggests that the language of users who disclose identity have some level of similarity regardless of identity type.

\section{Further details on propensity score matching}
\label{sec:sub:matching}
For each separate identity within the subcategory level, we use all covariates to train a logistic regression model using \texttt{scikit-learn}, which we use for assigning propensity scores to each sample. We stratify the scores into $N$ strata where $N$ equals the root number of positive samples. We use the Fisher Jenks natural break algorithm~\citep{jenks1967data} to obtain the strata bins, which we use for binning both \treated and \untreated users according to propensity scores. Within each strata, we assign matched pairs for each \treated user from the pool of \untreated users with the following steps. We first limit to \untreated users who changed their profiles in the same week as the \treated user. Next, we computed the Euclidean distance between the z-score normalized covariates to select up to 5 users with the shortest distance to the \treated user.

\begin{figure}[t!]
    \centering
    \includegraphics[width=0.7\textwidth]{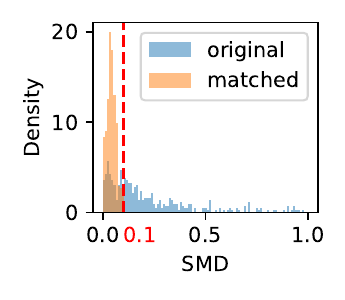}
    \caption{Distribution of standardized mean differences for all covariates between treated and control users (1) before and (2) after matching. The red line indicates 0.1, which most of the covariates fall into after matching. This indicates that the matching process significantly reduces potential confounders due to unbalanced covariates.}
    \label{fig:smd}
\end{figure}

\begin{figure}[t!]
    \centering
    \includegraphics[width=.7\textwidth]{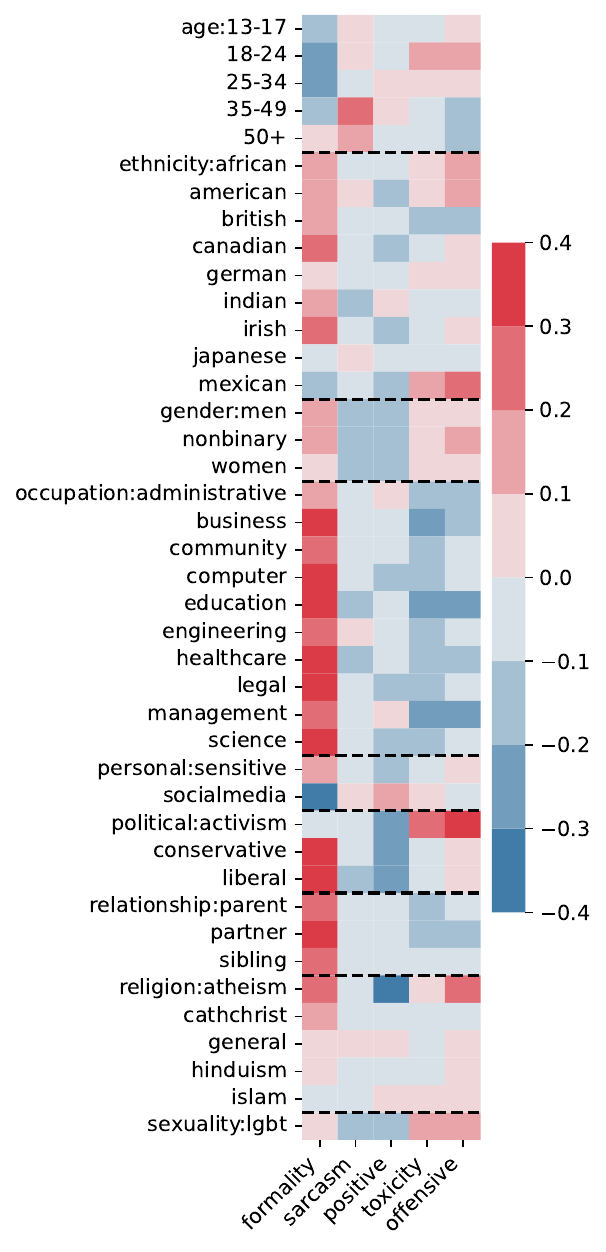}
    \caption{A heatmap comparing the correlations of identity-specific language with different styles. Similar categories exhibit similar styles.}
    \label{fig:style}
\end{figure}

\section{Measuring topic and stylistic distances between \treated and \alwayspositive users}
\label{sec:sub:style}
To measure topic distributions, for each identity we run zero-shot contextualized topic models~\citep{bianchi2021topic,bianchi2021zero} on the tweets of \alwayspositive users with 50 topics for 20 epochs, then obtain a 50-dimensional distribution which represents their topics $D^T_{AP}$. We then infer the topic distributions of the pre- and post-treatment tweets from \treated as $D^T_{pre}$ and $D^T_{post}$, which we use to measure the Jensen-Shannon distances of each distribution to $D^T_{AP}$. For style, we select five style variables from \citet{kang2021style} as well as classifier models from the Hugging Face API trained on public datasets: offensiveness~\citep{barbieri2020tweeteval}, formality~\citep{rao2018formality,pavlick2016formality}, sarcasm~\citep{misra2023Sarcasm}, toxicity~\citep{jigsaw2017toxicity,jigsaw2019toxicity}, and positive sentiment~\citep{hartmann2023sentiment}. For each identity, we computed the binary style scores for every tweet of the \alwayspositive users to obtain a $N\times5$ dimension matrix of style scores with $N$ as the number of tweets. We fitted PCA on the matrix to obtain the projection of its principal component, $D^S_{AP}$, which we use to represent the stylistic distribution of \alwayspositive users. Likewise, we obtained the same matrices for tweets from pre- and post-treatment periods of \treated users, and transformed these matrices into a single dimension using the principal component from fitted PCA of \alwayspositive, resulting in $D^S_{pre}$ and $D^S_{post}$. We then used cohen's d~\citep{cohen2013statistical} to compute the difference between each of the style distributions to $D^S_{AP}$.

\begin{figure}[t!]
    \centering
    \includegraphics[width=0.7 \textwidth]{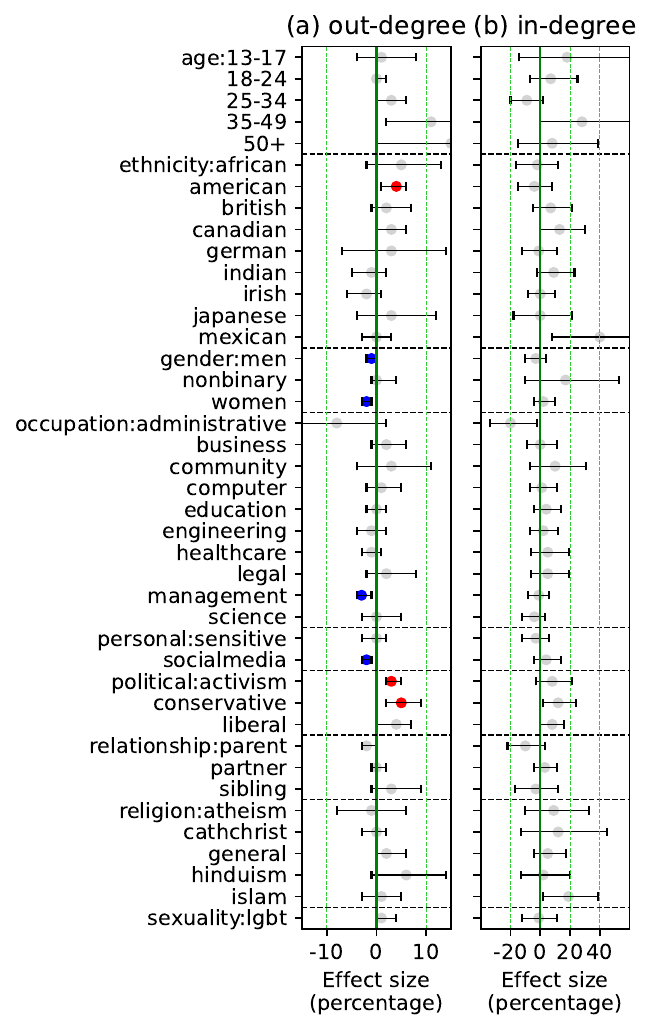}
    \caption{Plot on effect sizes of total network in- and out-degree regardless of identity type, where we see that the overall network does not increase as much, confirming the existence of intentional rewiring towards those of the same identity.}
    \label{fig:net_all}
\end{figure}

\begin{figure}[t!]
    \centering
    \includegraphics[width=0.7 \textwidth]{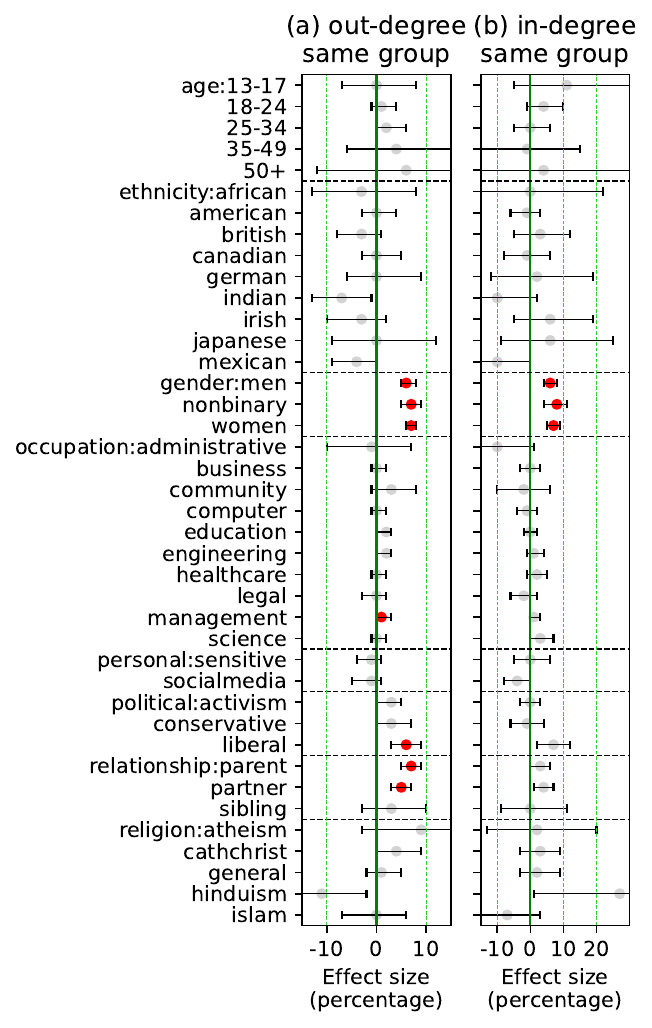}
    \caption{Plot on effect sizes of total network in- and out-degree when restricted to different identities in the same category. We observe increased connectivity between users who disclose gender and relationship status, especially parent and partner relationships.}
    \label{fig:net_other}
\end{figure}

\begin{figure}[]
    \centering
    \includegraphics[width=0.7 \textwidth]{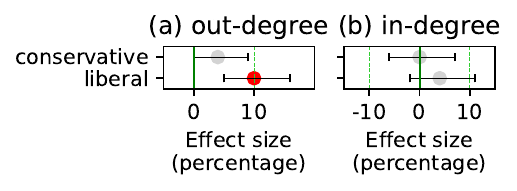}
    \caption{Plot on effect sizes of in- and out-degree when restricted to the opposite political ideology. Interestingly, we observe increased out-degree connectivity from those who declare themselves as liberals towards conservative users, but not from conservatives.}
    \label{fig:net_pol}
\end{figure}

\end{document}